\documentclass[12pt]{article}
\usepackage{amssymb}
\usepackage{mathrsfs}
\usepackage{bm}
\usepackage{graphicx}
\usepackage{subfigure}
\usepackage{hyperref}
\usepackage[utf8]{inputenc}
\begin{document}
%\begin{titlepage}
%\begin{flushright}
%v2.0\par\noindent
%\today
%\end{flushright}
\begin{center}
\begin{Large}
{\bf  Probing the holomorphic anomaly of the $D=2, \mathcal{N}=2$, Wess--Zumino model on the lattice}
\end{Large}
\vskip1truecm
Stam Nicolis\footnote{E-Mail: Stam.Nicolis@lmpt.univ-tours.fr}

{\sl CNRS--Laboratoire de Math\'ematiques et Physique Th\'eorique (UMR 7350)\\
F\'ed\'eration de Recherche ``Denis Poisson'' (FR 2964)\\
D\'epartement de Physique\\
Universit\'e ``Fran\c{c}ois Rabelais'' de Tours\\
Parc Grandmont, Tours 37200, France}

\end{center}

\vskip1truecm

\begin{abstract}
We study a generalization of the Langevin equation, that describes fluctuations, of commuting degrees of freedom, for scalar  field theories with  worldvolumes of arbitrary dimension, following Parisi and Sourlas and correspondingly generalizes the Nicolai map.  Supersymmetry appears inevitably, as  defining the consistent closure of system+fluctuations  and it can be probed by the identities satisfied by the correlation functions of the noise fields, sampled by the action of the commuting fields.  This can be done effectively, through numerical simulations.  

We focus on the case where the target space is invariant under global rotations, in Euclidian signature, corresponding to global Lorentz transformations, in Lorentzian signature. This can describe target space supersymmetry. 

In this case a cross--term, that is a total derivative for abelian isometries, or when the fields are holomorphic functions of their arguments, can lead to obstructions. We study its effects and find that, in two dimensions, it cannot lead to the appearance of the holomorphic anomaly, in any event, when fluctuations are taken into account,  because continuous symmetries can't be broken in two dimensions.

\end{abstract}
%\end{titlepage}
%\newpage
\section{Introduction}\label{intro}
The Langevin equation
\begin{equation}
\label{langevin}
\frac{d\phi}{d\tau} = -\frac{\partial W(\phi)}{\partial\phi(\tau)}+\eta(\tau)
\end{equation}
 describes the {\em equilibrium} fluctuations of a physical system, defined  by the, commuting, field(s) $\phi(\tau)$, in contact with a ``bath'', defined by the, commuting, field(s)  $\eta(\tau)$. The properties of the field(s) $\eta(\tau)$ are assumed known and the Langevin equation provides the map that allows the computation of the correlation functions, $\langle\phi(\tau_1)\cdots\phi(\tau_n)\rangle$, given the correlation functions, $\langle\eta(\tau_1)\cdots\eta(\tau_n)\rangle$. The mapping expresses the fact that the system is consistently closed and takes into account the ``backreaction'' of the system on the bath, if the bath is assumed to be described by a normalizable partition function, that can be taken to be a universal--non-zero--constant, whose value  can be set to 1, by a choice of units.
 
 The typical case is that of white noise:
 \begin{equation}
 \label{white_noise}
 \begin{array}{l}
 \displaystyle
 \left\langle\eta(\tau)\right\rangle = 0\\
 \displaystyle
 \left\langle\eta(\tau)\eta(\tau')\right\rangle = \nu\delta(\tau-\tau')
 \end{array}
 \end{equation}
 with the other correlation functions deduced from Wick's theorem.  Other noise distributions are, of course, possible--``colored noise'', in particular, where the 2--point function isn't ultra--local, has been studied for many applications--though many issues pertaining to it remain to be elucidated and are the topic of current research. 
 
 It should be stressed that, depending on the physical meaning ascribed to the parameter, $\nu$, this framework describes quantum fluctuations (if $\nu=\hbar$), thermal fluctuations (if $\nu=k_\mathrm{B}T$) or fluctuations due to disorder, if $\nu$ is the strength of the disorder. 
 
 In the above expressions, $\tau$ labels the worldvolume, that's, therefore, one--dimensional. This means, that there's just one ``bath'', in which the physical system of interest is immersed.
  
 The partition function of the noise, that describes these correlation functions,  is given by the expression
 \begin{equation}
 \label{Znoise}
 Z_\mathrm{L} = \int\,[{\mathscr D}\eta(\tau)]\,e^{-\int\,d\tau\,\frac{1}{2}\eta(\tau)^2}=1
 \end{equation} 
 and the corresponding partition function for the field(s) $\phi(\tau)$ by the expression, through a change of variables
 \begin{equation}
 \label{Zphi}
 1 = Z_L = \int\,[{\mathscr D}\phi(\tau)]\,\left|\mathrm{det}\left(\frac{\delta\eta(\tau')}{\delta\phi(\tau)}\right)\right|\,e^{-\int\,d\tau\,\frac{1}{2}\left(\frac{d\phi}{d\tau}+W'(\phi)\right)^2}=\left\langle\left|\mathrm{det}\left(\frac{\delta\eta(\tau')}{\delta\phi(\tau)}\right)\right|\right\rangle_\mathrm{QM}Z_\mathrm{QM}
 \end{equation}
 where we have introduced the subscript ``L'' to highlight that $Z_\mathrm{L}$ is deduced from the Langevin equation, while the subscript ``QM'' highlights that the partition function $Z_\mathrm{QM}$ is the (Euclidian) partition function we write down when studying the physical system in the ``canonical'' ensemble.  We shall use terminology appropriate to statistical mechanics and Euclidian quantum field theory.  These expressions define what it means that the physical system is in equilibrium with the fluctuations of the ``bath'' and, therefore, that the system + fluctuations is consistently closed. These statements rely on the properties of the potential--that it is bounded from below and confines sufficiently rapidly at infinity for the partition function(s) to exist and the corresponding correlation functions to be well--defined. These conditions are necessary for lattice formulations to have a chance of leading to scaling limits. How to describe the case of a potential that describes scattering states remains to be elucidated.

 We would like to explore a generalization to higher--dimensional worldvolumes.  The reason this is interesting is that this is how  field theories, that involve an infinite (and, typically, an indefinite) number of particles,  can be described. A particularly interesting subclass are relativistic field theories, since it's known~\cite{leutwyler} that it isn't possible to describe a fixed number of particles in interaction in a way that is consistent with Lorentz invariance (integrable theories, that are the, apparent, exception, are, of course, free theories in disguise: the interaction, in this case, is a coordinate artifact in the space of fields.) 
 
 One such approach consists in replacing  eq.~(\ref{langevin}) by the following expression
 \begin{equation}
 \label{worldvolume_langevin}
 \sigma_A^{IJ}\frac{\partial\phi_J}{\partial u_A} = -\frac{\partial W}{\partial\phi_I}+\eta^I(u)
 \end{equation} 
 In this expression $u_A$, with $A=1,2,\ldots,d$, are the coordinates of the worldvolume and the noise field, $\eta(u)$, is taken to describe a white noise process, with correlation functions given by eq.~(\ref{white_noise}) and partition function by eq.~(\ref{Znoise}), with $\tau$ replaced by the $u_A$. When we replace $\eta(u)$ in the corresponding expression for the partition function, we remark that, if we take the $\sigma_A$ as the generators of the Clifford algebra, 
 \begin{equation}
\label{clifford}
\left\{\sigma_A,\sigma_B\right\}=2\delta_{AB}
\end{equation}
 where we're working in Euclidian signature, then we are, naturally, led to studying theories with as many scalar fields, $\phi(u)$, as the size of the representation of the $\sigma$ matrices. The simplest case is that of $d=2$, with the $\sigma$ as the Pauli matrices--then two scalar fields are required and there are two noise fields, as well. If $d=4$, then the corresponding matrices would be the Dirac $\gamma-$matrices; however this representation is reducible and can be expressed in terms of the Pauli matrices, also--the issues that arise are, whether the fermions can be taken to be Majorana, Weyl or both; and the answer depends on the spacetime dimension and the signature, Lorentzian or Euclidian. We shall work in Euclidian signature.
  
The expression of the LHS of eq.~(\ref{worldvolume_langevin}) is driven by well--defined problems in mind, but is sufficiently flexible for leading to interesting computational strategies, whatever the specific problem. 
 
The chief motivation for choosing that the $\sigma_A$ satisfy eq.~(\ref{clifford}) is because, in this way, the worldvolume is invariant under (global) rotations, in Euclidian signature--that can be identified with Lorentz transformations, in Minkowski signature--and the fields transform in the appropriate representations, also.

 This was studied in ref.~\cite{parisi_sourlas}, but there is much more work that remains to be done, following their insights--and later work hasn't elaborated on whether taking into account the fermions in the indirect way proposed by Parisi and Sourlas was, indeed, feasible.  What distracted from it was its identification with the dimensional reduction, that was quite quickly found to be of limited validity. The object of the present contribution is to show that the original idea of Parisi and Sourlas can, indeed,  be realized in a computationally well--defined way for studying the lattice actions of fermionic fields, on flat worldvolumes of arbitrary dimension, in terms of the lattice actions of their bosonic superpartners in an exact way, thereby extending the work on quantum mechanics, the theory of fields on one--dimensional worldvolumes~\cite{nicolis}. What is important to keep in mind is that the backreaction of the physical degrees of freedom on the bath can't be neglected.  
  
In ref.~\cite{nicolis} we showed by numerical simulations that  the correlation functions of the scalar, computed with $Z_\mathrm{L}$ and $Z_\mathrm{QM}$, were the same and equal to the correlation functions of the $d=1$, maximally supersymmetric Wess--Zumino model. Here we wish to consider an infinite number of scalars, coupled as per eq.~(\ref{worldvolume_langevin}). The question we would like to address is, whether, in this case, also, it's possible to describe $Z_\mathrm{L}$ and the corresponding $Z_\mathrm{SUSY}$ by the partition function, $Z_\mathrm{QM}$, of the theory of scalar fields, if we study the correlation functions of the noise field(s), that define the generalization of the  Nicolai map~\cite{nicolai}
\begin{equation}
\label{nicolai_map}
\eta^I(u)=\sigma_A^{IJ}\frac{\partial\phi_J(u)}{\partial u^A} + \frac{\partial W(\phi)}{\partial\phi_I(u)}
\end{equation}
In particular, we would like to understand  how the backreaction of the scalar(s) on the bath is realized.

For the absolute value of the Jacobian can be written as follows:
\begin{equation}
\label{jacobian_fermions}
\begin{array}{l}
\displaystyle
\left|\mathrm{det}\frac{\delta\eta(\tau)}{\delta\phi(\tau')}\right|=\mathrm{sign}\left( \mathrm{det}\left( \left[\sigma_A\right]_{IJ}\frac{\partial}{\partial u_A}+ \frac{\partial^2 W(\phi)}{\partial\phi^I\partial\phi^J}\right)\right)\times\\
\displaystyle
\hskip2truecm
\int\,[{\mathscr D}\psi(\tau)][{\mathscr D}\chi(\tau)]\,e^{\int\,d^d u\,\psi^I\left([\sigma_A]_{IJ}\frac{\partial}{\partial u_A}+
\frac{\partial^2 W}{\partial\phi^I\partial\phi^J}\right)\chi^J}
\end{array}
\end{equation} 
So, formally, $Z_\mathrm{L}$ can be written as
\begin{equation}
\label{Zsusy}
\begin{array}{l}
\displaystyle
Z_\mathrm{L}=\int\,[{\mathscr D}\phi][{\mathscr D}\psi][{\mathscr D}\chi]\,\mathrm{sign}\left( \mathrm{det}\left( \left[\sigma_A\right]_{IJ}\frac{\partial}{\partial u_A}+ \frac{\partial^2 W(\phi)}{\partial\phi^I\partial\phi^J}\right)\right)\times \\
\displaystyle
e^{-\int\,d^du\,\left\{\frac{1}{2}\left(\sigma_A\frac{\partial\phi}{\partial u_A}+W'(\phi) \right)^2-\psi\left(\sigma_A\frac{\partial}{\partial u_A}+W''(\phi)\right)\chi\right\}} = \\
\displaystyle
\left\langle \mathrm{sign}\left( \mathrm{det}\left( \left[\sigma_A\right]_{IJ}\frac{\partial}{\partial u_A}+ \frac{\partial^2 W(\phi)}{\partial\phi^I\partial\phi^J}\right)\right)\right\rangle_\mathrm{SUSY} Z_\mathrm{SUSY}
\end{array}
\end{equation}
where the partition function $Z_\mathrm{SUSY}$ defines a classical action, that's invariant under supersymmetric transformations: transformations that map the commuting fields to the anticommuting fields (and vice versa) and the  anticommutator  of two such transformations closes on the generator of the (Euclidian) translations. 

What is a non--trivial statement about how supersymmetry can be broken this way is that only the 1--point function of the noise field becomes non--zero under the change of variables; the 2--point function remains ultra--local. How anomalies might affect this was studied, in the mean field approximation, in ref.~\cite{nicolis_zerkak}.

Of course the same argument applies in the other direction: the fluctuations of a bosonic theory are consistently described by the dynamics of the fermionic superpartners. It's just that it's much easier, computationally, to work with commuting rather than anticommuting fields. So the important point is that supersymmetry, as  discussed here, isn't a feature to be added by hand--it's a, well--hidden, property of known theories.

In the following sections we shall recall the relevant properties of the ``minimal'' $d=2$ Wess--Zumino model, that's described by such a generalized Langevin equation and focus on its bosonic part and the noise fields. We shall study the case of the cubic superpotential, for which the holomorphic anomaly is, classically, absent, if the boundary conditions are chosen appropriately, since the cross term is, then, a total derivative; as well as the case of the quartic superpotential, for which holomorphic factorization, classically, breaks down, as well as more general cubic superpotentials, that lead to the same scalar potential.   By computing the identities, satisfied by the noise field, however, we shall provide evidence that the holomorphic anomaly is, indeed, absent. This is the result of straightforward Monte Carlo simulations of the bosonic action. We discuss how boundary conditions can affect these statements.
We end with our conclusions and ideas of directions of further inquiry, namely to gauge theories and to more than two spacetime  dimensions.

\section{The two--dimensional Wess--Zumino model}\label{WZ2}
The action of the ``minimal'' $d=2$ Wess--Zumino model, defined by the generalized Langevin equation, when the worldvolume is two--dimensional, describes a theory with ${\mathcal N}=2$ supersymmetry~\cite{parisi_sourlas}. The classical action is given by the expression 
\begin{equation}
\label{WZ2classical}
\begin{array}{l}
\displaystyle
S=\int\,d^2x\,\left\{\frac{1}{2}\left( \sigma_A\partial_A\phi+W'(\phi)\right)^2-\psi(x)\sigma_A\partial_A\chi(x)-\psi(x)W''(\phi)\chi(x)  \right\}=\\
\displaystyle
\int\,d^2x\,\left\{ \frac{1}{2}\left(\partial_A\phi_J\right)^2 + \sigma_A^{IJ}\partial_A\phi_J\frac{\partial W}{\partial\phi_I} + \frac{1}{2}\frac{\partial W}{\partial\phi_I}\frac{\partial W}{\partial\phi_I} -\psi\sigma_A\partial_A\chi-\psi W''(\phi)\chi\right\}=\\
\displaystyle
\int\,d^2x\,\left\{ \frac{1}{2}\left(\partial_A\phi_J\right)^2 -\frac{F_I^2}{2} +  F_I\frac{\partial W}{\partial\phi_I} + \sigma_A^{IJ}\partial_A\phi_J\frac{\partial W}{\partial\phi_I}
-\psi\sigma_A\partial_A\chi-\psi W''(\phi)\chi\right\}
\end{array}
\end{equation}
This expression is, simply, the result of expanding the noise fields  and what we wish to check is how their fluctuations are constrained by this fact.

There is a term, that we, apparently, could have  omitted, were we working with worldvolumes, invariant only under translations (and were using periodic boundary conditions):
\begin{equation}
\label{surfaceterm}
\sigma_A^{IJ}\int\,d^2x\,\frac{\partial\phi_J}{\partial u_A}\frac{\partial W}{\partial\phi_I}
\end{equation}
This would be a total derivative, were $\sigma_A^{IJ}\propto\delta^{IJ}$, i.e. were the representation of the algebra, generated by the $\sigma_A$ matrices,  one--dimensional; the worldvolume would then be invariant under an abelian isometry group. (It will, also, be a total derivative, for the linear terms in the superpotential.)

 For worldvolumes that are invariant under non--abelian isometry groups,  as is the case here, for the $\sigma_A$, that generate an $SU(2)$ algebra, this term isn't, manifestly, a total derivative. So, in general,  we must keep it. 
 
It is this term that relates the bosons to the fermions, in the bosonic action, through the appearance of the matrices $\sigma_A$ and might be considered a sort of spin--orbit coupling. 

 For one--dimensional worldvolumes this  term is, manifestly, a total derivative (however, it can describe anomalies~\cite{cecotti_girardello} in this way); for higher dimensional worldvolumes, that have richer isometry groups, more care is required, in order to understand its role for the dynamics, and one aim of this paper is to try and define computationally efficient strategies for describing its consequences. 
 
So we would like to devise ways of testing, whether this term is, indeed, present, or is, in fact,  cancelled by some mechanism.

There are two ways to resolve this issue: show that the contribution of this term is cancelled by the contribution of the fermionic determinant; or show that there are cases where this term is, in fact, a surface term, for dynamical reasons. 

 There are two cases, that look very similar, that illustrate the difference.
 
For instance, if the equations that define the superpotential are given by the expressions
\begin{equation}
\label{cubicsuperpot_totder}
\begin{array}{l}
\displaystyle
\frac{\partial W}{\partial\phi_1}=g\left(\phi_1^2-\phi_2^2\right)\\
\displaystyle
\frac{\partial W}{\partial\phi_2}=2g\phi_1\phi_2
\end{array}
\end{equation} 
then a straightforward calculation shows that the crossterm is a sum of total derivatives: 
\begin{equation}
\label{crossterm_cubicplus}
\sigma_A^{IJ}\frac{\partial\phi_J}{\partial u_A}\frac{\partial W}{\partial\phi_I}=\frac{\partial}{\partial x}\left( \phi_1^2\phi_2 -\frac{\phi_2^3}{3}\right) +
\frac{\partial}{\partial y}\left(\frac{\phi_1^3}{3}-\phi_1\phi_2^2 \right)
\end{equation}
without having to impose the Cauchy--Riemann equations on the scalars. 

It's interesting to note that  the, seemingly, very similar equations
\begin{equation}
\label{cubicsuperpot_nototder}
\begin{array}{l}
\displaystyle
\frac{\partial W}{\partial\phi_1}=g\left(\phi_1^2-\phi_2^2\right)\\
\displaystyle
\frac{\partial W}{\partial\phi_2}=-2g\phi_1\phi_2
\end{array}
\end{equation} 
that do define a holomorphic function, 
\begin{equation}
\label{Wcubicholo}
W(\phi_1,\phi_2) = \frac{g}{6}\left(\phi_1+\mathrm{i}\phi_2\right)^3 + \frac{g}{6}\left(\phi_1-\mathrm{i}\phi_2\right)^3 = \frac{1}{6}\left(\Phi^3 + \overline{\Phi}^3\right)=
\frac{\phi_1^3}{3}-\phi_1\phi_2^2
\end{equation}
do not imply that the crossterm is a total derivative, unless the scalars satisfy the Cauchy--Riemann equations. 

Both superpotentials lead to the same scalar potential, 
\begin{equation}
\label{scalarpotcubic}
V(\phi_1,\phi_2)=
\frac{1}{2}\left[
\left(\frac{\partial W}{\partial\phi_1}\right)^2 + 
\left(\frac{\partial W}{\partial\phi_2}\right)^2\right]=
\frac{g^2}{2}\left[\left(\phi_1^2 + \phi_2^2\right)^2\right]^2=\frac{g^2}{2}|\Phi|^4
\end{equation}
The question this fact raises, is, whether the two systems are, in fact,  different, or not.  What might distinguish them is the behavior of the noise fields. That's what we shall focus on in the numerical study. 

In both cases the scalars are massless. 

Another example, where the cross term isn't a surface term, unless the scalars are holomorphic, too, is provided by the quartic superpotential, 
\begin{equation}
\label{quartic_superpot}
W(\phi_1,\phi_2)=\frac{g}{4!}\left(\phi_1^2+\phi_2^2-v^2\right)^2=\frac{g}{4!}\left(\Phi\overline{\Phi}-v^2\right)^2
\end{equation}
(even when $v^2=0$) that is allowed, by power counting, in two (and three) spacetime dimensions, since the scalar potential is of sixth degree in the field. The corresponding scalar potential is given by the expression 
\begin{equation}
\label{Vscalar_quarticsup}
V=\frac{1}{2}\left\{ \left(\frac{\partial W}{\partial\phi_1}\right)^2 + \left(\frac{\partial W}{\partial\phi_2}\right)^2  \right\}=
\frac{g^2}{72}\left(\phi_1^2+\phi_2^2-v^2\right)^2\left(\phi_1^2+\phi_2^2\right)
\end{equation}

 This is, therefore, another test case for whether the holomorphic anomaly is, indeed, cancelled by the fluctuations. For this potential, when $v^2=0$, the crossterm is given by the expression
\begin{equation}
\label{quartic_crossterm}
\sigma_A^{IJ}\frac{\partial\phi_J}{\partial u_A}\frac{\partial W}{\partial\phi_I} = \frac{g}{6}\left(\phi_1^2+\phi_2^2\right)\left\{  \frac{\partial}{\partial x}\left(\phi_1\phi_2\right) + \frac{1}{2}\frac{\partial}{\partial y}\left( \phi_1^2-\phi_2^2\right) \right\}
\end{equation}
The fact that the prefactor isn't a constant is the obstruction to it being a total derivative; unless the fields are holomorphic functions of their arguments, i.e. massless. However, since, in two dimensions, continuous symmetries can't be spontaneously broken, it is expected that this term doesn't contribute in any event,  since the expectation value of the prefactor does vanish. And, indeed, the lattice potential possesses a global minimum at the origin and local minima, that are  never degenerate with it, but break the global symmetry, away from it. These can't survive, when fluctuations are taken into account. The details of the numerical study will be presented in forthcoming work.

\begin{figure}[thp]
\caption[]{Typical results for the cubic superpotential: $\langle\eta_n^I\eta_{n+d}^J\rangle$ for $I=J$ (left panel) and $I\neq J$ (right panel) and $d=0,2,4,6,8$, on the $17\times 17$ square lattice. $g_\mathrm{latt}^2=0.7$. The diagonal noise term is a $\delta-$function, while the off-diagonal noise term vanishes, to numerical precision.}
\subfigure{\includegraphics[scale=0.4]{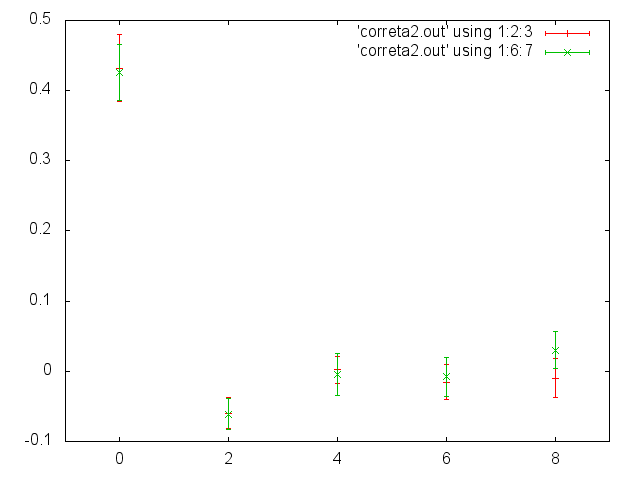}}
\subfigure{\includegraphics[scale=0.4]{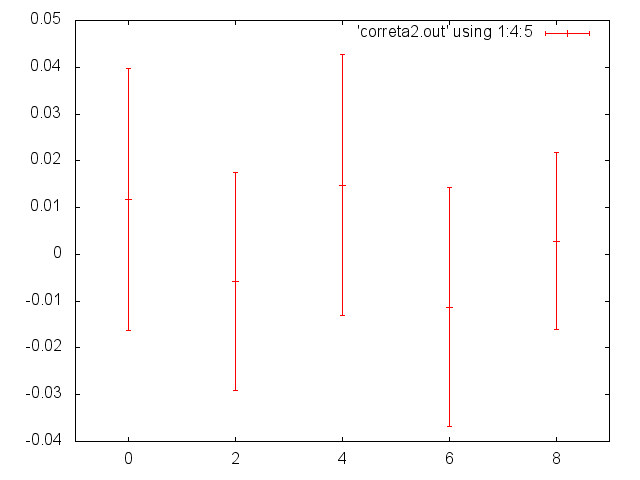}}
\label{cubicfig}
\end{figure}

We present the 1-- and 2--point functions for the noise fields, for the cubic  superpotential, in figs.~\ref{cubicfig} In both cases we note that the global SO(2) symmetry is manifest, since $\langle\eta^1\eta^1\rangle=\langle\eta^2\eta^2\rangle$.  Details, including the case of the quartic superpotential,  will be presented in forthcoming publications. 
\section{Conclusions}\label{conclusions}
We have generalized the consistent description of the separation of dynamical and bath degrees of freedom by the Langevin equation to worldvolumes with non--abelian isometries. It is in this way that it's possible to describe the fluctuations in terms of fields, rather than particles. Imposing appropriate conditions on the coefficients  of the worldvolume derivatives, namely that they generate a Clifford algebra, it's possible to describe target space fermions, that are related to the, commuting, dynamical degrees of freedom, by target space supersymmetry, that, once more, describes the consistent closure of the system.  In this way it's possible to recover the results obtained by Parisi and Sourlas and generalize them; and show that the holomorphic anomaly, that appears at the classical level, is eliminated by the quantum fluctuations, that forbid the breaking of global continuous symmetries, in two dimensions. 

The calculation of the fluctuations of the action of the bosonic part of the action, inevitably, takes into account the backreaction of the fermionic part; and vice versa. It is thus possible to  generalize the approach, studied for one--dimensional worldvolumes in ref.~\cite{nicolis} to worldvolumes that induce target space Lorentz invariance and supersymmetry. 

It would, also, be interesting to understand the dynamics of multiplets that, apparently, don't contain any auxiliary fields (e.g. the (4,4,0) multiplet)~\cite{fedoruk_et_al,Baulieu:2016tkh,Kuznetsova:2013tf}) within the stochastic framework. This seems to imply  studying multiplicative noise for the Langevin equation, that's, also, relevant for the stochastic dynamics of magnets~\cite{NTT}, where the formulation in  curved space becomes necessary~\cite{Kozyrev:2017kiy}.

{\bf Acknowledgements}: It's a pleasure to thank S. A. Fedoruk, E. A. Ivanov and A. O. Sutulin for the wonderful workshop SQS2017 at Dubna, that fostered, once more, stimulating exchanges. I would, also, like to acknowledge discussions with M. Axenides, E. Floratos and J. Iliopoulos. 

\newpage

\bibliographystyle{utphys}
\bibliography{Dubna17bib}

\providecommand{\href}[2]{#2}\begingroup\raggedright\begin{thebibliography}{10}

\bibitem{leutwyler}
H.~Leutwyler, ``A no-interaction theorem in classical relativistic hamiltonian
  particle mechanics,'' \href{http://dx.doi.org/10.1007/BF02749856}{{\em Il
  Nuovo Cimento (1955-1965)} {\bfseries 37} no.~2, (May, 1965) 556--567}.
  \url{https://doi.org/10.1007/BF02749856}.

\bibitem{parisi_sourlas}
G.~Parisi and N.~Sourlas, ``{Supersymmetric Field Theories and Stochastic
  Differential Equations},''
\href{http://dx.doi.org/10.1016/0550-3213(82)90538-7}{{\em Nucl. Phys.}
  {\bfseries B206} (1982) 321--332}.
%%CITATION = NUPHA,B206,321;%%.

\bibitem{nicolis}
S.~Nicolis, ``{How quantum mechanics probes superspace},''
  \href{http://arxiv.org/abs/1606.08284}{{\ttfamily arXiv:1606.08284
  [hep-th]}}.
[Phys. Part. Nucl. Lett.14,no.2,357(2017)].
%%CITATION = ARXIV:1606.08284;%%.

\bibitem{nicolai}
H.~Nicolai, ``{Supersymmetry and Functional Integration Measures},''
\href{http://dx.doi.org/10.1016/0550-3213(80)90460-5}{{\em Nucl. Phys.}
  {\bfseries B176} (1980) 419--428}.
%%CITATION = NUPHA,B176,419;%%.

\bibitem{nicolis_zerkak}
S.~Nicolis and A.~Zerkak, ``{Supersymmetric probability distributions},''
  \href{http://dx.doi.org/10.1088/1751-8113/46/28/285401}{{\em J. Phys.}
  {\bfseries A46} (2013) 285401},
\href{http://arxiv.org/abs/1302.2361}{{\ttfamily arXiv:1302.2361 [hep-th]}}.
%%CITATION = ARXIV:1302.2361;%%.

\bibitem{cecotti_girardello}
S.~Cecotti and L.~Girardello, ``{Stochastic Processes in Lattice (Extended)
  Supersymmetry},''
\href{http://dx.doi.org/10.1016/0550-3213(83)90200-6}{{\em Nucl. Phys.}
  {\bfseries B226} (1983) 417--428}.
%%CITATION = NUPHA,B226,417;%%.

\bibitem{fedoruk_et_al}
S.~Fedoruk, E.~Ivanov, and A.~Smilga, ``{$\mathcal N=$ 4 mechanics with diverse
  (4, 4, 0) multiplets: Explicit examples of hyper-Kähler with torsion,
  Clifford Kähler with torsion, and octonionic Kähler with torsion
  geometries},'' \href{http://dx.doi.org/10.1063/1.4871440}{{\em J. Math.
  Phys.} {\bfseries 55} (2014) 052302},
\href{http://arxiv.org/abs/1309.7253}{{\ttfamily arXiv:1309.7253 [hep-th]}}.
%%CITATION = ARXIV:1309.7253;%%.

\bibitem{Baulieu:2016tkh}
L.~Baulieu and F.~Toppan, ``{Chains of topological oscillators with instantons
  and calculable topological observables in topological quantum mechanics},''
  \href{http://dx.doi.org/10.1016/j.nuclphysb.2016.05.030}{{\em Nucl. Phys.}
  {\bfseries B912} (2016) 88--102},
\href{http://arxiv.org/abs/1610.00943}{{\ttfamily arXiv:1610.00943 [hep-th]}}.
%%CITATION = ARXIV:1610.00943;%%.

\bibitem{Kuznetsova:2013tf}
Z.~Kuznetsova and F.~Toppan, ``{Effects of Twisted Noncommutativity in
  Multi-particle Hamiltonians},''
  \href{http://dx.doi.org/10.1140/epjc/s10052-013-2483-x}{{\em Eur. Phys. J.}
  {\bfseries C73} (2013) 2483},
\href{http://arxiv.org/abs/1301.5501}{{\ttfamily arXiv:1301.5501 [hep-th]}}.
%%CITATION = ARXIV:1301.5501;%%.

\bibitem{NTT}
S.~Nicolis, P.~Thibaudeau, and J.~Tranchida, ``{Finite-dimensional colored
  fluctuation-dissipation theorem for spin systems},''
  \href{http://dx.doi.org/10.1063/1.4975132}{{\em AIP Adv.} {\bfseries 7}
  (2017) 056012},
\href{http://arxiv.org/abs/1610.01622}{{\ttfamily arXiv:1610.01622
  [cond-mat.stat-mech]}}.
%%CITATION = ARXIV:1610.01622;%%.

\bibitem{Kozyrev:2017kiy}
N.~Kozyrev, S.~Krivonos, O.~Lechtenfeld, A.~Nersessian, and A.~Sutulin,
  ``{${\cal N}{=}\,4$ supersymmetric mechanics on curved spaces},''
\href{http://arxiv.org/abs/1711.08734}{{\ttfamily arXiv:1711.08734 [hep-th]}}.
%%CITATION = ARXIV:1711.08734;%%.

\end{thebibliography}\endgroup

\end{document}